\documentclass[12pt]{spieman}  % 12pt font required by SPIE;
\usepackage{amsmath,amsfonts,amssymb}
\usepackage{graphicx}
\usepackage{setspace}
\usepackage{tocloft}
\usepackage{lineno}
\usepackage{booktabs}
\title{Dimpled scalar vortex coronagraph laboratory demonstration}

\author[a*]{Niyati Desai}
\author[a]{Garreth Ruane}
\author[a]{Susan Redmond}
\author[a,b]{Dimitri Mawet}
\author[a]{Eugene Serabyn}
\author[a]{Bertrand Mennesson}
\affil[a]{Jet Propulsion Laboratory, California Institute of Technology, Pasadena, California, United States}
\affil[b]{Department of Astronomy at California Institute of Technology, Pasadena, California, United States}

\cftpagenumbersoff{figure}
\cftpagenumbersoff{table} 
\begin{document} 
\maketitle

\begin{abstract}
Achieving the Habitable Worlds Observatory (HWO) goal of $10^{-10}$ contrast at a separation of $\sim3\lambda/D$ across a $\ge$20\% bandwidth requires coronagraph focal plane masks with both broadband high contrast performance and high planet throughput. Scalar vortex coronagraphs (SVCs) offer a promising alternative to polarization-sensitive vector vortex designs but face chromatic limitations. This work presents the latest laboratory demonstrations of second-generation scalar vortex prototypes that incorporate radial phase dimples to improve broadband starlight suppression. We compare these new “dimpled” sawtooth masks to previous-generation scalar designs through high-contrast imaging experiments on the In-Air Coronagraph Testbed. Using electric field conjugation, we achieve near testbed-limited contrasts across both narrow (2\%) and broadband (10\%) spectral ranges. We report the best in-air contrasts achieved to date for scalar vortex masks across narrow and broadband spectral ranges and we also show that the dimpled vortex predicted bench-limited contrast performances for 2\%, 10\% and 18\% bandwidths agree with the measured lab contrasts within a factor of two. These results highlight the potential of topographically achromatized scalar vortex masks as candidates for future space-based high-contrast imaging missions and mark a significant step toward polarization-independent coronagraphs capable of meeting HWO performance requirements. 
\end{abstract}

% Include a list of up to six keywords after the abstract

% Include email contact information for corresponding author
{\noindent \footnotesize\textbf{*}Address all correspondence Niyati Desai:  \linkable{ndesai@jpl.nasa.gov} }

\begin{spacing}{1}   % use double spacing for rest of manuscript

\section{Introduction}
\label{sect:intro}  % \label{} allows reference to this section
% Exoplanet direct imaging context, goals for HWO, need for high performing FPMs
% Scalar versus vector vortex coronagraph, limitations of each
% Major contributions of this work:  latest tented results of new scalar vortex prototypes achieving the limit of in-air coronagraph contrast performance across broadband

Directly imaging and characterizing Earth-like exoplanets around sun-like stars is one of the primary science goals driving the design of NASA’s Habitable Worlds Observatory (HWO). To detect biosignatures in reflected light, HWO must achieve starlight suppression at the $10^{-10}$ contrast level across broad spectral bandwidths ($\ge$20\%), while maintaining high throughput for small angular separations ($\sim3 \lambda/D$). Meeting these stringent requirements places exceptional demands on the coronagraph focal plane mask (FPM), which must suppress starlight efficiently across wavelength and polarization\cite{Mennesson2024}.

The two approaches to vortex-based coronagraphy are the vector vortex coronagraph (VVC)\cite{Mawet2005} and the scalar vortex coronagraph (SVC)\cite{Swartzlander2006}. VVCs employ polarization-dependent geometric phase, which enables broadband performance but imparts phase ramps in opposite directions for left and right handed polarization states. They are limited by polarization leakage caused by manufacturing deviations from the desired 180$^\circ$ retardance. In practice they often require filtering to isolate only one of the polarization states and results in 50\% reduced throughput. On the other hand, SVCs rely on a longitudinal phase delay imparted through thickness variation or refractive index modulation effectively creating spatially varying optical path differences. They can achieve theoretically polarization-independent operation at the expense of chromaticity\cite{Swartzlander2006,Ruane2019}. Overcoming this chromatic limitation is essential for establishing scalar vortex designs as a viable option for HWO’s broadband, polarization-independent coronagraph.

This work presents high contrast laboratory demonstrations of next-generation scalar vortex coronagraphs incorporating radial phase dimples to mitigate chromatic leakage and enhance broadband starlight suppression. Newly developed scalar vortex masks need to be evaluated on high-contrast imaging testbeds and benchmarked directly against existing coronagraphic designs to quantify their performance gains. Demonstrating these scalar vortex coronagraphs in realistic laboratory conditions is essential to advancing their overall technological maturity for consideration in future space-based high-contrast imaging missions.

\section{A Review of Scalar Vortex Coronagraphs}
% History of SVC development, from first theory (Foo, Swartzlander papers), and simulation (Ruane 2019) to new designs over the past 5 years
% Reviewing methods of achromatization including metasurfaces (Palatnick and Koenig) and topography variations (Desai and Galicher)

The vortex coronagraph suppresses on-axis starlight by introducing an azimuthally varying phase ramp, $e^{il\theta}$, onto the focal plane wavefront\cite{Mawet2005,Foo2005}, where $l$ is the ``topological charge" or the number of $2\pi$ phase wraps around the center. When propagated to the pupil plane, this helical phase redistributes starlight outside the geometric pupil, where it can be blocked by a Lyot stop, enabling deep contrast at small inner working angles. The scalar implementation achieves this phase modulation through longitudinal optical path differences—typically by varying the thickness of a transmissive dielectric material or refractive index across the focal plane mask\cite{Swartzlander2006, Ruane2019}.

While the SVC offers polarization-independent operation and high throughput, its performance is inherently chromatic, as the phase delay scales with wavelength. Achieving broadband starlight suppression therefore requires achromatization strategies that maintain the desired vortex phase ramp across a wide spectral range.

Over the past two decades, several approaches have been developed to mitigate this chromaticity, including metasurfaces and surface modulation. Metasurface implementations use subwavelength features to engineer a broadband phase delay\cite{konig2023,Konig24,Konig25,Palatnick2025}. Surface modulation optimizes the mask’s physical shape to approximate the ideal phase ramp over a range of wavelengths\cite{Galicher2020,Desai_2022}.

Among these approaches, topography-based scalar vortex designs have demonstrated the most promising laboratory performance. Wrapped staircase and sawtooth masks have reached broadband contrasts on the order of $10^{-7}$ from 3--10 $\lambda/D$ on an in-air coronagraph testbed, limited by their chromatic topographies\cite{Desai_2023}. New topographies have recently been proposed which reduce chromatic leakage by integrating a central phase dimple similar to the original Roddier and Roddier phase coronagraph\cite{Roddier_1997,Desai_2024Roddier}. This topography solution was found by describing the behavior of a scalar vortex through a modal decomposition of its purely azimuthal phase. In this framework, the power in the 0th mode corresponds to zeroth-order leakage—light that passes straight through the vortex without acquiring any vortex phase and is therefore not suppressed—which arises only off the central wavelength. To counteract this effect, a central $2\pi$ phase dimple is introduced, which phase-shifts roughly half of the total PSF energy concentrated within a small central radius, effectively canceling the leakage term. Further details on the theory of this design, the optimization of this dimple radius jointly with the scalar vortex, and on the modal decomposition can be found in Desai et al. 2024 JATIS\cite{Desai_2024Roddier}. In this work we experimentally validate this previously proposed approach, comparing the broadband performance of new ``dimpled'' scalar vortex prototypes against previous-generation purely azimuthal scalar vortex masks.

\section{Dimpled Vortex Design, Manufacturing and Testing}

\begin{figure}
    \centering
    \includegraphics[width=0.8\linewidth]{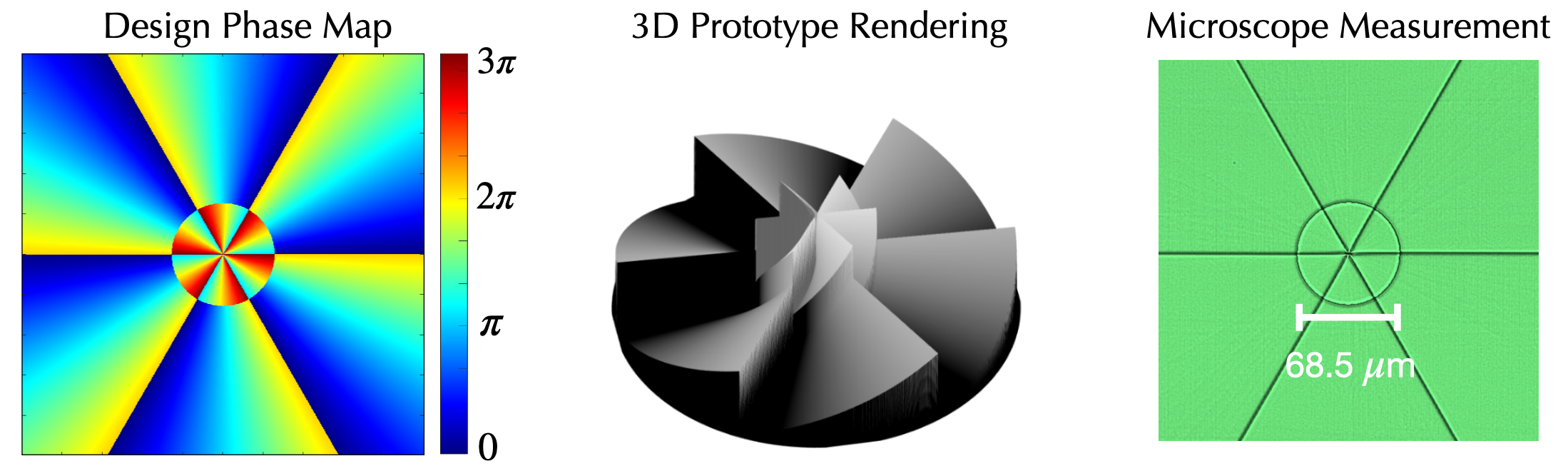}
    \caption{(left) Two-dimensional map of the desired spatially-varying phase shift showing central $\pi$ shifted region. (middle) Three-dimensional schematic of dimpled scalar vortex design. (right) A microscope image of the mask manufactured in fused silica.}
  \label{fig:mask_images}
\end{figure}

This work focuses on the experimental demonstration of the new dimpled scalar vortex topography in a high contrast coronagraph testbed. The primary goal is to evaluate the chromatic leakage mitigation of this design, which is done by measuring the broadband starlight suppression. The new dimpled vortex combines an azimuthal sawtooth vortex with a central region that is $\pi$ phase shifted, forming an azimuthal and radial phase mask optimized to reduce chromatic leakage across visible wavelengths. Figure~\ref{fig:mask_images} shows the design phase map for the dimpled scalar vortex, the 3D rendering of this design for the etched prototype, and the microscope image of the mask whose results are reported in this paper. 

Several scalar masks with this dimpled vortex design were fabricated using fused silica substrates and the same high-precision direct laser writing and reactive ion etching (RIE) processes used in earlier generation devices.\cite{Desai_2023} Details on similar earlier prototype manufacturing can be found in Desai et al. 2024 SPIE Proceedings\cite{Desai_2024SPIE}. Each mask measured 15$\times$15$\times$1~mm , with the vortex phase pattern confined to a 14$\times$14~mm central region. A broadband anti-reflective coating was applied to both sides, yielding reflectance below 0.5\% across 585--835~nm. Masks were designed to have varying central wavelengths between 655~nm and 685~nm and for an F/\#~=~83, corresponding to the operating bands and parameters of JPL's in-air testbed at the focal plane mask location\cite{Baxter2021}. For optimal broadband performance across a 20\% band, the dimple was designed to have a radius of 0.621~$\lambda/D$ (as found to achieve best contrast for in combination with the sawtooth vortex in Desai et al. 2024 JATIS\cite{Desai_2024Roddier}) which translates to a diameter of $\sim$69~microns. Microscope measurements of the mask, shown on the right of Figure~\ref{fig:mask_images}, found the dimple to be approximately 68.5 microns, well within manufacturing tolerances for this design according to previous parameter sensitivity analysis of the dimple height and size\cite{desai2023_Roddier,Desai_2024Roddier}.

One mask prototype with a measured etch height corresponding to $\lambda_0 = 658$~nm was integrated into the In-Air Coronagraph Testbed (IACT) at JPL for high-contrast testing\cite{Baxter2021}. The IACT is an off-axis reflective bench operating at visible wavelengths with a pupil diameter of 7.6~mm and a polarized tunable broadband light source. The testbed has a single DM (34$\times$34 actuator kilo-DM from Boston Micromachines) and implements pairwise probing (PWP) and electric field conjugation (EFC) for wavefront correction. Previous works explored PWP and other EFC methods with both scalar and vector vortex masks and found consistent contrast floor measurements for this in-air testbed to be near $10^{-8}$\cite{Desai_2024SCC,desai2025}.

Closed-loop wavefront control was performed using pairwise probing and electric field conjugation (EFC)\cite{Riggs2018}. Each experiment began with a 2\% narrowband correction to establish baseline performance and alignment, followed by 10\% EFC with 5 subbands and 18\% broadband testing to assess chromatic stability. The central wavelength $\lambda_0$ for testing was determined by using Zygo measurements provided by the vendor and translating the measured etch height to the corresponding wavelength for 0--2$\pi$ phase coverage. Prior model-mismatch investigations guided the model parametrization of the mask clocking angle and central wavelength for EFC, both identified as critical factors for broadband EFC convergence\cite{desai2025}.

\section{Results}

\begin{figure}
    \centering
    \includegraphics[width=1\linewidth]{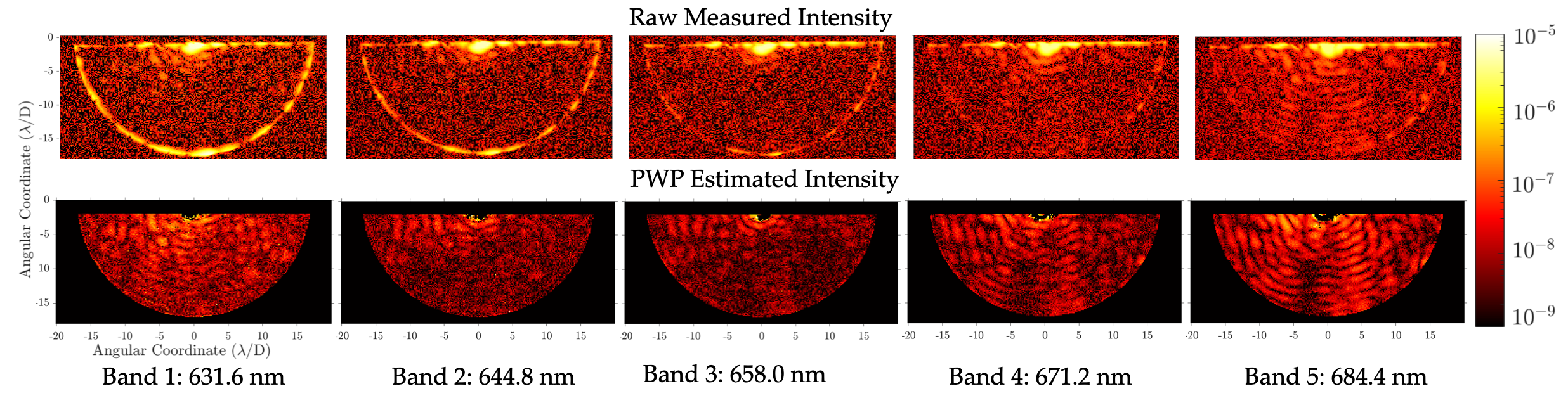}
    \caption{Electric field conjugation (EFC) was performed across a 10\% band with 5 subbands shown here. The estimated and measured intensity match closely and the residual speckles are more visible at offset wavelengths.
}
    \label{fig:pwp_subbands}
\end{figure}

The dimpled scalar vortex prototypes were evaluated on IACT to quantify their starlight suppression performance relative to previous-generation sawtooth designs. A half dark hole was generated from 3--18~$\lambda/D$, matching the size of the physical field stop on IACT, and scored from 3--10~$\lambda/D$, initially in narrowband (2\%), and then in broadband (10\%) using five subbands each with 2\% bandwidths. The normalized intensity is then computed as the ratio of residual starlight to the peak intensity for the non-coronagraphic point spread function and is reported throughout this paper as the `average contrast'. Figure~\ref{fig:pwp_subbands} shows the five subbands centered around $\lambda_0$~=~658~nm which were used during the 10\% bandwidth PWP and dark hole digging process on IACT. The close agreement of the estimated and measured intensity images indicates a well-matched EFC model to the experimental setup. Additionally the chromatic speckles are clearly more visible at the edges of the band than at the central wavelength ($\lambda_0$).

\begin{figure} [t]
    \centering
    \includegraphics[width=\linewidth]{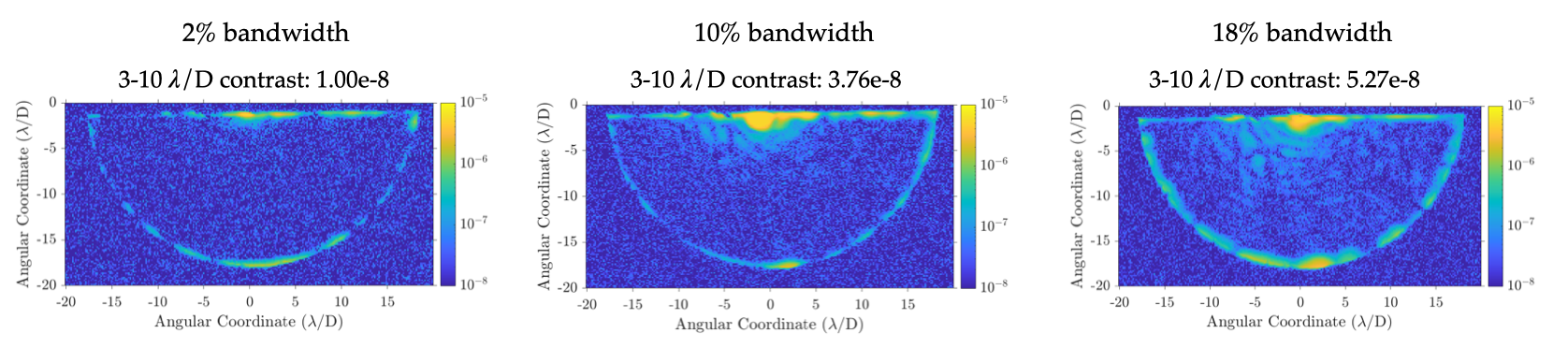}
    \caption{In-air coronagraphic testbed dark holes (from left to right): narrowband (2\%), broadband (10\%), and broadband (18\%). Dark holes were dug from 3 - 18 $\lambda/D$ and scored between 3 - 10 $\lambda/D$.}
    \label{fig:dhs}
\end{figure}

By using IACT's SuperK Varia tunable filter, after the 10\% broadband EFC was performed, the chromatic performance was further evaluated at wavelengths across a larger 18\% bandwidth for model comparison. The final broadband dark holes in Figure~\ref{fig:dhs} show each of these three bandwidths. The left image shows the 2\% narrowband dark hole achieved with the dimpled sawtooth prototype at 658~nm. The half dark hole resulted in an average contrast of 1.00e-8 between 3--10~$\lambda/D$. This represents a factor of $\sim$2 improvement over the best result obtained with the first-generation sawtooth mask under identical conditions\cite{Desai_2023}. This narrowband contrast is consistent with the measured testbed contrast floor, indicating that the dimpled vortex mask is performing near the limit of IACT.

Broadband EFC also demonstrated significant improvements in chromatic performance relative to previous designs. The dimpled vortex achieved an average contrast of 3.76e-8 across the 10\% band, compared to 3.24e-7 for the earlier generation\cite{Desai_2023} for 3-10 $\lambda/D$, corresponding to an order of magnitude improvement. Furthermore with the same final DM shape, the measured contrast opening up the band further to 18\% is only slightly degraded to 5.27e-8 across 3-10 $\lambda/D$, less than a factor of two.

\begin{figure}
    \centering
    \includegraphics[width=\linewidth]{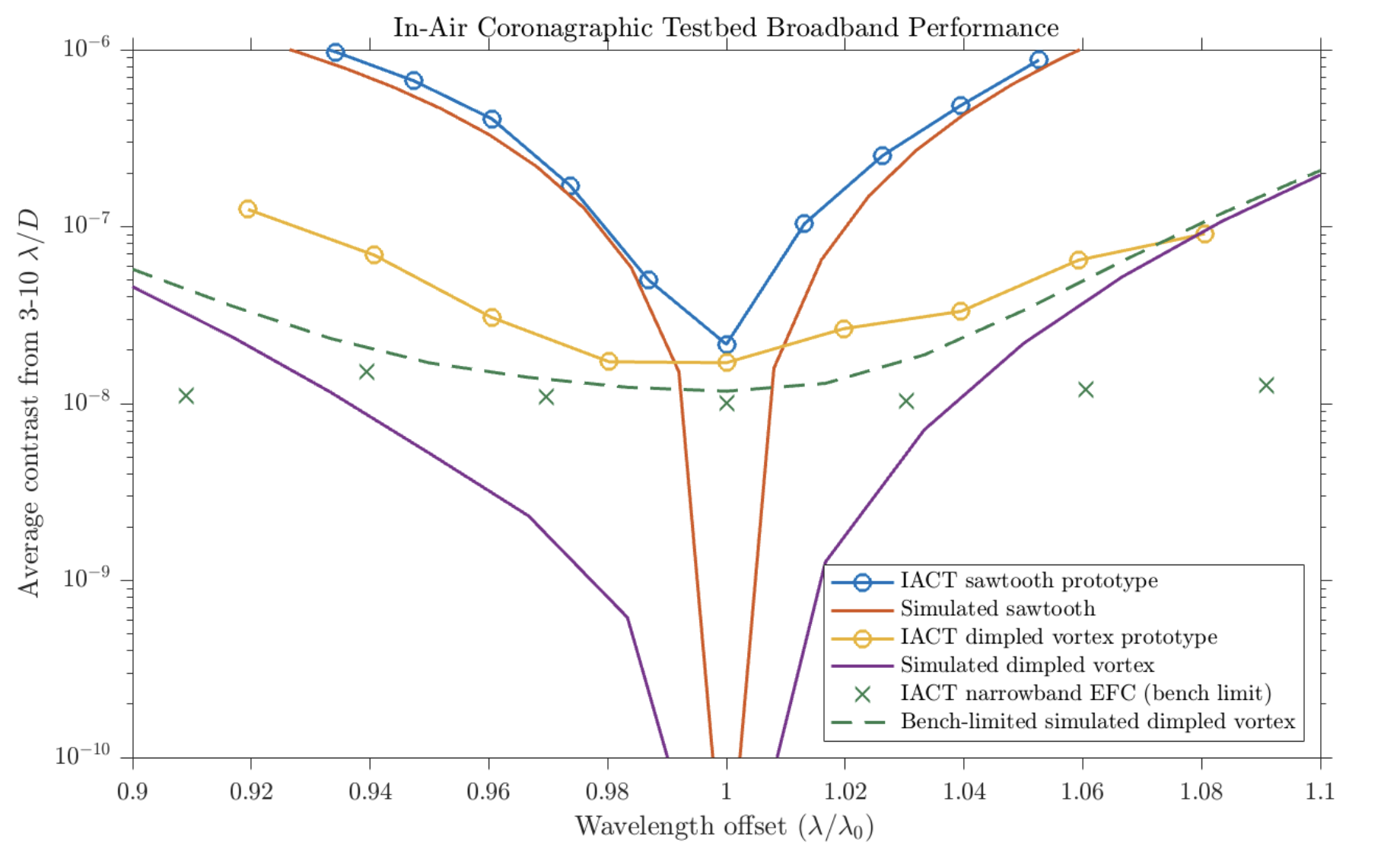}
    \caption{In-air coronagraphic testbed V-curve demonstrating contrast as a function of wavelength offset.}
    \label{fig:vcurve}
\end{figure}

Figure~\ref{fig:vcurve} compares the final contrast curves for narrowband and broadband performance, along with an estimate of the IACT bench limit derived from narrowband EFC measurements. The computed average contrasts from this plot for the simulated, lab measured, and expected bench-limited performances are summarized in Table~\ref{tab:contrasts}. The so-called ``V-curve analysis'' in Figure ~\ref{fig:vcurve} also shows the nearly overlapping match between the simulated and experimental contrast performance for the previous sawtooth vortex. Notably, the dimpled vortex performance over broader bandwidths (up to 20\%), is limited by the in-air testbed contrast floor. By averaging the narrowband EFC contrasts across the 20\% band (shown in Figure~\ref{fig:vcurve} as green crosses), we can estimate the flat average broadband IACT contrast floor between 600~nm and 720~nm to be approximately 1.17e-8. Adding this average broadband testbed limit estimate to the simulated dimpled vortex curve yields a predicted broadband testbed-limited contrast (shown in Figure~\ref{fig:vcurve} as a green dashed line). The average of this estimated curve for 18\% is 3.88e-8 and 1.85e-8 for 10\% bandwidth assuming a perfect wavefront with no aberrations and no DM correction. These are only a factor of $\sim$2 from the measured average broadband contrasts shown in Figure~\ref{fig:dhs}. The remaining mismatch between the experimentally measured (shown in yellow) and expected bench-limited contrast curves for the dimpled vortex is the asymmetry in the predicted curve which is not seen in the experimental data. This can be attributed to the fact that broadband EFC solutions equally weight all the wavelengths across the full bandwidth and tend to find an optimal DM shape which yield a symmetric V-curve.

Although the experimental IACT data is not perfectly overlapping the expected bench-limited performance curve for the dimpled vortex like for the sawtooth vortex, this data shows a significant improvement of the new design over the previous SVC performance. The 18\% bandwidth contrast curve shows the expected gradual degradation in contrast and maintains suppression within a factor of three of the 10\% result. The broadband results here are limited by the in-air bench performance and need to be further evaluated in conditions where the limit of the bench is below the predicted simulated performance. However these results already successfully confirm that radial dimples mitigate chromatic leakage, enabling the scalar vortex to approach the current limit of IACT.

{\renewcommand{\arraystretch}{1.6}
\begin{table} [t!]
\begin{center}
\resizebox{\textwidth}{!}{
    \begin{tabular}{rccccc}
    \toprule
        % Bandwidth & PWP+EFC & SCC+EFC & iEFC\\ 
        & \multicolumn{2}{c}{Sawtooth Scalar Vortex} & \multicolumn{3}{c}{Dimpled Scalar Vortex} \\
        % \cmidrule(r){2-3} \cmidrule(r){4-5} 
         & 2\% & 10\% & 2\% & 10\% & 18\% \\
         % \cmidrule(r){2-2} \cmidrule(r){3-3} \cmidrule(r){4-4} \cmidrule(r){5-5} 
        \cmidrule(r){2-3} \cmidrule(r){4-6}
         Simulated performance* & $1.21 \times 10^{-8}$ & $3.24 \times 10^{-7}$ & $6.35 \times 10^{-11}$ & $6.73 \times 10^{-9}$ & $2.70 \times 10^{-8}$\\ 
         Measured on IACT & $2.16 \times 10^{-8}$ & $3.37 \times 10^{-7}$ & $1.00  \times 10^{-8}$ & $3.76 \times 10^{-8}$ & $5.27  \times 10^{-8}$\\
         Bench-limited model & $1.17 \times 10^{-8}$ & $3.36 \times 10^{-7}$ & $1.17 \times 10^{-8}$ & $1.85 \times 10^{-8}$ & $3.88 \times 10^{-8}$\\

         \bottomrule
         \multicolumn{6}{p{1.2\textwidth}}{*Narrowband vortex simulations are model resolution limited. Resolution parameters were chosen so the monochromatic theoretically perfect null is at least below 1e-10.}\\
    \end{tabular}
    }
        \vspace{5 mm} 
    \caption{\label{tab:contrasts} Summary table of average simulated and measured contrasts for 2\%, 10\%, and 18\% bandwidths for the previous sawtooth scalar vortex and new dimpled scalar vortex.}
\end{center}
\end{table}
}

\section{Discussion and Conclusion}
% Need to go into vacuum next!!
% - these results are with only a single DM! no amplitude correction!
% Also new designs needed to reach HWO $10^{-10}$ levels

Overall, these results demonstrate that dimpled scalar vortex coronagraphs can achieve near in-air contrast limits across both narrow and broadband spectral ranges, marking a significant step toward polarization-independent coronagraphs suitable for future space missions. These results confirm that radial phase dimples effectively mitigate chromatic leakage, addressing a key limitation of earlier scalar vortex coronagraph designs and establishing a new performance benchmark for scalar vortex coronagraphs.

We demonstrated record broadband performance of scalar vortex coronagraphs using dimpled vortex designs, achieving average contrasts of 3.76e-8 in 10\% bandwidth and 5.27e-8 in 18\% bandwidths from 3-10 $\lambda/D$ with only a single DM on an in-air coronagraphic testbed. This demonstrated contrast level is already sufficient for next-generation ground-based observatories, including coronagraph instruments planned for facilities such as TMT and the ELT.

Further improvements in contrast are likely possible with an additional DM allowing for not only phase correction but also chromatic control and amplitude correction. Since these results demonstrate that the dimpled vortex achieves in-air bench-limited performance, they motivate the need for testing these masks in vacuum and on testbeds with deeper contrast floors. The next steps in scalar vortex testing must focus on removing testbed limitations and validating true broadband performance. Further improvements in fabrication accuracy, clocking agreement with the model, and model-based control will also be critical for scaling these results to higher contrast levels.

To reach HWO’s $10^{-10}$ contrast goal, new scalar vortex designs that build upon these results will be necessary. This type of study can be easily expanded to other topographies, higher vortex charges, or alternate techniques of achromatizing scalar vortex phase masks. Some avenues that are currently being explored include moving to a multi-material approach\cite{Ruane2019,Swartzlander2006}, developing a scalar metasurface implementation\cite{Palatnick2025,konig2023,Konig25}, and combining the scalar vortex with several gratings\cite{Konig2025SPIE,Errmann2013ABS}. With demonstrated agreement between modeled designs and experimental lab measurements, this work establishes a clear path toward scalar vortex coronagraphs as a viable, polarization-independent pathway for future space-based high-contrast imaging.

\subsection{Acknowledgments}
The research was carried out at the Jet Propulsion Laboratory, California Institute of Technology, under a contract with the National Aeronautics and Space Administration (80NM0018D0004). N.D. is supported by an appointment to the NASA Postdoctoral Program at the Jet Propulsion Laboratory, California Institute of Technology, administered by Oak Ridge Associated Universities under contract with NASA.

%  Acknowledgments and funding information should be added after the conclusion, and before references. The acknowledgments section does not have a section number. Include grant numbers and the full name of the funding body. Use of large language models and other AI tools for language and grammar clean-up should also be disclosed here.

\subsection{Disclosures}
The authors declare that there are no financial interests, commercial affiliations, or other potential conflicts of interest that could have influenced the objectivity of this research or the writing of this paper.

\subsection{Code and Data Availability}
All data in support of the findings of this paper are available within the article and simulation code can be freely accessed on the FALCO github repository (https://github.com/ajeldorado/falco-matlab/tree/SVC)\cite{Riggs2018}.
%%%%% References %%%%%

\bibliography{report}   % bibliography data in report.bib
\bibliographystyle{spiejour}   % makes bibtex use spiejour.bst

\end{spacing}
\end{document}